\documentclass[%
 reprint,
 amsmath,amssymb,
 aps,
]{revtex4-1}

\usepackage{graphicx}
\usepackage{dcolumn}
\usepackage{bm}
\usepackage{braket}
\usepackage{graphicx} 
\usepackage{color}                                                                                             

\newcommand{\beginsupplement}{%
        \setcounter{table}{0}
        \renewcommand{\thetable}{S\arabic{table}}%
        \setcounter{figure}{0}
        \renewcommand{\thefigure}{S\arabic{figure}}%
     }
\newcommand{\ex}[1]{\mathrm{e}^{#1}}

\newcommand{\pa}[1]{\left(#1 \right)}

\newcommand{\BR}[1]{\Biggl[#1 \Biggr]}

\newcommand{\ca}[1]{\mathcal{#1}}

\newcommand{\abs}[1]{\left|#1\right|}

\newcommand{\ar}[1]{\xrightarrow[#1]{}}

\newcommand{\fr}{\frac}
\newcommand{\s}[1]{\sqrt{#1}}

\def\be{\begin{equation}}
\def\ee{\end{equation}}
\def\ba{\begin{eqnarray}}
\def\ea{\end{eqnarray}}

 \def\d{{\delta}}

 \def\ba{{\bar{\alpha}}}

 \def\b{{\beta}}
 \def\e{{\epsilon}}

\makeatletter
\let\cat@comma@active\@empty
\makeatother

\begin{document}


\title{Derivation of holographic negativity in AdS$_3$/CFT$_2$ }
\author{Yuya Kusuki}
\email{yuya.kusuki@yukawa.kyoto-u.ac.jp}
\affiliation{Center for Gravitational Physics,
Yukawa Institute for Theoretical Physics (YITP), Kyoto University,
Kitashirakawa Oiwakecho, Sakyo-ku, Kyoto 606-8502, Japan.}

\author{Jonah Kudler-Flam}%
\email{jkudlerflam@uchicago.edu}
\author{Shinsei Ryu}
\email{ryuu@uchicago.edu}
\affiliation{%
 Kadanoff Center for Theoretical Physics, University of Chicago, Chicago, Illinois~60637, USA
}%

\date{\today}

\begin{abstract}
We present a derivation of the holographic dual of logarithmic negativity in AdS$_3$/CFT$_2$ that was recently conjectured in [Phys. Rev. D 99, 106014 (2019)]. This is given by the area of an extremal cosmic brane that terminates on the boundary of the entanglement wedge. The derivation consists of relating the recently introduced R\'enyi reflected entropy to the logarithmic negativity in holographic conformal field theories. Furthermore, we clarify previously mysterious aspects of negativity at large central charge seen in conformal blocks and comment on generalizations to generic dimensions, dynamical settings, and quantum corrections.
\end{abstract}

\maketitle



\textit{Introduction.---}The von Neumann entropy of the reduced density matrix is an excellent measure of the entanglement between bipartite subsystems in a pure state. In particular, it has played a major role in the understanding of how bulk geometry holographically emerges from microscopic degrees of freedom in the AdS/CFT correspondence. This is due to the fact that the von Neumann entropy of a boundary subregion $A$ is equal to the area of the extremal bulk surface $\gamma_A$ that is homologous to $A$, \cite{2006PhRvL..96r1602R,2006JHEP...08..045R,2007JHEP...07..062H}
\begin{align}
    S_{vN}(\rho_A) = \frac{\mbox{area($\gamma_A$)}}{4 G_N},
\end{align}
where $G_N$ is the bulk Newton constant. For mixed states $\rho_{AB}$, the von Neumann entropy fails to serve as a correlation measure because $S_{vN}(\rho_A) \neq S_{vN}(\rho_B)$ and the corresponding mutual information [$I_{AB}\equiv S_{vN}(\rho_A)+S_{vN}(\rho_B)-S_{vN}(\rho_{AB}) $] fails to quantify the entanglement between $A$ and $B$. Rather, the mutual information captures classical correlations, such as thermodynamic entropy, on top of the quantum correlations. 

In general, one may be interested in characterizing the entanglement structure of many-body systems in mixed states. In particular, it is an interesting question to ask if and how mixed state entanglement manifests itself geometrically in the bulk in AdS/CFT. For this purpose, we study the logarithmic negativity, a suitable measure of entanglement for mixed states \footnote{It is important to note that logarithmic negativity only provides an upper bound on the distillable entanglement.}, based on the positive partial transpose criterion \cite{PhysRevLett.77.1413,1996PhLA..223....1H,1999JMOp...46..145E,2000PhRvL..84.2726S,2002PhRvA..65c2314V,2005PhRvL..95i0503P}. While most mixed state entanglement measures are defined in terms of intractable optimization procedures, the logarithmic negativity is operationally defined and generally computable.
For a bipartite density matrix $\rho_{AB}$, the partial transpose is an operation that transposes just one of the subsystems
\begin{align}
    \bra{i_A ,j_B} \rho_{AB}^{T_B} \ket{k_A, l_B} = \bra{i_A ,l_B} \rho_{AB} \ket{k_A, j_B},
\end{align}
where $i_A, j_B,k_A,$ and $l_B$ are bases for subsystems $A$ and $B$. The logarithmic negativity is then defined as
\begin{align}
    \mathcal{E}(\rho_{AB}) \equiv \log \left|\rho_{AB}^{T_B} \right|_1,
\end{align}
where $\left| \mathcal{O} \right|_1 = \mbox{Tr} \sqrt{\mathcal{O}\mathcal{O}^{\dagger}}$ is the trace norm. 


With motivations from quantum error-correcting codes and preliminary examples in 2D conformal field theory, two of the authors conjectured that logarithmic negativity in holographic conformal field theories is dual to a backreacted entanglement wedge cross section in asymptotically anti-de Sitter (AdS) space-times \cite{PhysRevD.99.106014}, giving a concrete proposal for how mixed state entanglement is geometrized in the bulk. In this Letter, we provide a proof of the conjecture. For symmetric configurations, we provide direct comparisons between conformal field theory (CFT) computations of negativity and entanglement wedge cross sections.

\textit{Holographic reflected entropy.---}Additional quantities have been shown to be related to the entanglement wedge cross section in holographic theories, including the entanglement of purification, odd entropy, and reflected entropy \cite{2018NatPh..14..573U,2018JHEP...01..098N,PhysRevLett.122.141601,2019arXiv190500577D}. While each quantity is intriguing in its own right, we will use the reflected entropy for our purposes in deriving the holographic dual for logarithmic negativity in AdS$_3$/CFT$_2$.

We start by reviewing the construction of the reflected entropy and then state the conjecture for the holographic dual of logarithmic negativity (\ref{holo_neg_conj}). 
Generically, mixed density matrices may be decomposed into a sum of pure states
\begin{align}
    \rho_{AB} = \sum_a p_a \rho_{AB}^{(a)}.
\end{align}
We may then perform a Schmidt decomposition on each pure state
\begin{align}
    \rho_{AB}^{(a)} = \sum_{i,j} \sqrt{\lambda_i^{(a)} \lambda_j^{(a)} }\ket{i^{(a)}}\bra{j^{(a)}}_A \otimes \ket{i^{(a)}}\bra{j^{(a)}}_B.
\end{align}
A canonical purification of the original mixed state may then be constructed in the doubled Hilbert space $\mathcal{H}_A \otimes \mathcal{H}_{A^*}\otimes \mathcal{H}_{B}\otimes \mathcal{H}_{B^*}$
\begin{align}
    \ket{\sqrt{\rho_{AB}}} \equiv \sum_{a,i,j} \sqrt{p^{(a)}\lambda_i^{(a)} \lambda_j^{(a)} }\ket{i^{(a)}}_A\ket{j^{(a)}}_{A^*}\ket{i^{(a)}}_B\ket{j^{(a)}}_{B^*}.
\end{align}
The reflected entropy is then defined by \cite{2019arXiv190500577D}
\begin{align}
    S_R(A:B) \equiv S_{vN}(\rho_{AA^*}),
\end{align}
where $S_{vN}$ is the von Neumann entropy and $\rho_{AA^*}$ is the reduced density matrix on $\mathcal{H}_A \otimes \mathcal{H}_{A^*}$. In holographic conformal field theories, this was shown to be dual to twice the area of the entanglement wedge cross section \cite{2019arXiv190500577D}
\begin{align}
    S_R = 2 E_W.
\end{align}
This was proven for time reflection symmetric states using the Lewkowycz-Maldacena gravitational replica trick \cite{2013JHEP...08..090L}. It is natural to think that such a correspondence holds in generic time-dependent settings. In fact, many nontrivial checks of the time-dependent conjecture have been performed in Ref.~\cite{2019arXiv190706646K}. 

We consider the R\'enyi reflected entropies
\begin{align}
    S_R^{(n)}(A:B) \equiv S^{(n)}(\rho_{AA^*}).
\end{align}
As shown by Dong \cite{2016NatCo...712472D}, the R\'enyi entropies are related to the modular entropies which are dual to cosmic branes in the bulk gravity theory
\begin{align}
    \tilde{S}^{(n)} \equiv n^2 \partial_n \left( \frac{n-1}{n}S^{(n)}\right) = \frac{\mbox{area(Cosmic Brane$_n$)}}{4G_N}.
\end{align}
The cosmic branes are codimension-two objects with tension
\begin{align}
    T_n = \frac{n-1}{4n G_N}.
\end{align}
The modular reflected entropies are then dual to twice the area of cosmic branes that terminate on the entanglement wedge \footnote{In higher dimensions, the Engelhardt-Wall gluing procedure may be complicated by shockwaves associated with the conical singularities \cite{2019JHEP...05..160E}.} 
\begin{align}
    \tilde{S}_R^{(n)} \equiv n^2 \partial_n \left( \frac{n-1}{n}S_R^{(n)}\right) =2 \frac{\mbox{area(Cosmic Brane$_n$)}}{4G_N}\Big|_{E_W}.
\end{align}
While this was not explicitly stated in Ref.~\cite{2019arXiv190500577D}, it is a simple corollary of their gravitational construction of the state $\ket{\sqrt{\rho_{AB}}}$.


\textit{Holographic negativity conjecture.---}In Ref.\ \cite{PhysRevD.99.106014}, it was conjectured that the logarithmic negativity in holographic conformal field theories is dual to a backreacting entanglement wedge cross section. The conjecture may be concisely stated in terms of the R\'enyi reflected entropy as
\begin{align}
    \mathcal{E} = \frac{S_R^{(1/2)}}{2}
    \label{holo_neg_conj}
\end{align}
for holographic conformal field theories. In the special cases where both the subregion configurations and the states are spherically symmetric, this backreaction may be accounted for by \footnote{Strictly speaking, this equality is only proven for the vacuum state \cite{2011JHEP...12..047H,2014JHEP...10..060R}. However, it is expected (and has passed multiple tests) that it holds whenever both the subsystem configuration and quantum state are spherically symmetric. We stress that it does not hold in generic situations.}
\begin{align}
    \mathcal{E} = \mathcal{X}_d E_W,
    \label{simple_LN}
\end{align}
where $\mathcal{X}_d$ is a constant that depends on the dimension of the CFT
\begin{align}
    \mathcal{X}_d &= \left(\frac{1}{2}x_d^{d-2}\left(1 + x_d^2\right) - 1\right), \\ x_d &= \frac{2}{d}\left(1 + \sqrt{1 - \frac{d}{2} + \frac{d^2}{4}}\right).
\end{align}
We will consider 2D CFTs where $\mathcal{X}_2 = 3/2$. 
We will explicitly compute the holographic negativity for disjoint intervals in the vacuum state and a single interval at finite temperature from conformal blocks, finding precise agreement with (\ref{simple_LN}). 

\textit{Deriving holographic negativity.---}We now provide a simple derivation of (\ref{holo_neg_conj}) in 2D CFTs. We consider two arbitrary subsystems
\begin{align}
    A = \bigcup\limits_{i=1}^{n_A} \left[u_i, v_i \right], \quad B = \bigcup\limits_{i=1}^{n_B} \left[w_i, y_i \right].
\end{align}
The logarithmic negativity in the vacuum state (and generic conformal transformations from the vacuum) may be computed by a correlation function of twist fields
\begin{align}
    \mathcal{E} = \lim_{n_e \rightarrow 1} \log \left<\prod_{i}^{n_A}\left( \sigma_{n_e}(u_i) \bar{\sigma}_{n_e}(v_i)\right)\prod_{i}^{n_B}\left( \bar{\sigma}_{n_e}(w_i) \sigma_{n_e}(y_i)\right) \right>.
    \label{LN_corr}
\end{align}
See Ref.~\cite{2013JSMTE..02..008C} for a thorough exposition of computing negativity in conformal field theory.
The twist fields have conformal dimensions 
\begin{align}
    h_{n_e} &= \bar{h}_{n_e} = \frac{c}{24}\left( n_e-\frac{1}{n_e}\right), 
    \\
    h^{(2)}_{n_e} &= \bar{h}^{(2)}_{n_e} = \frac{c}{12}\left(\frac{n_e}{2}-\frac{2}{n_e} \right),
\end{align}
where $h^{(2)}_{n_e}$ are the double twist fields that arise when fusing two twist fields of the same chirality.
We compare \eqref{LN_corr}
to the correlation function of generalized twist fields that computes half of the R\'enyi reflected entropy
\begin{align}
    &S_R^{(1/2)}/2 =\lim_{m\rightarrow 1} \lim_{n\rightarrow 1/2} \log \Bigg<\prod_{i}^{n_A}\left( \sigma_{g^{\ }_A}(u_i) \sigma_{g_A^{-1}}(v_i)\right) \nonumber
    \\
    &\quad 
    \times
    \prod_{i}^{n_B}\left( {\sigma}_{g^{\ }_B}(w_i) \sigma_{g_B^{-1}}(y_i)\right)
 \Bigg>_{{\it CFT}^{\otimes mn}}.
 \label{SR_corr}
\end{align}
These generalized twist fields have the action of moving fields between sheets in two directions labeled by $m$ and $n$. The $g_B g_A^{-1}$ twist field does not appear in the correlation function, rather in the conformal block, as it is the lowest weight primary operator in the operator product expansion (OPE) of $g_B$ and $g_A^{-1}$. For more precise definitions, see Ref.~\cite{2019arXiv190500577D}. The generalized twist fields have conformal dimensions
\begin{align}
    &
    h_{g^{\ }_B} = h_{g_A^{-1}} = \frac{cn(m^2-1)}{24m}, 
    \\ 
    &
    h_{g^{\ }_Bg_A^{-1}} = \frac{2c(n^2-1)}{24n}.
\end{align}
The conformal dimensions, positions, and dominant intermediate channels of the operators in (\ref{LN_corr}) and (\ref{SR_corr}) precisely match. Thus, in the limit of large central charge, we confirm (\ref{holo_neg_conj}) for the class of states that may be obtained by conformal transformations from the vacuum. For completely generic states that include primary operator insertions, we are also able to confirm (\ref{holo_neg_conj}), though we leave the details to the Supplemental Material.


\textit{Symmetric examples.---}
In symmetric configurations, we may use the simplification of (\ref{simple_LN}).
First, we focus on the negativity of disjoint intervals in the vacuum. 
The negativity is a conformally invariant quantity \cite{2012PhRvL.109m0502C} and computed by

\begin{equation}
\ca{E}=\lim_{n_e \to 1} \log\braket{ \sigma_{n_e}(\infty) \bar{\sigma}_{n_e}(1) \bar{\sigma}_{n_e}(x,\bar{x}) \sigma_{n_e}(0)   },
\label{LN_disjoint_corr}
\end{equation}
where we set the two intervals to $A= [0,x ]$ and $B=[ 1, \infty ]$ for simplicity.
We restrict ourselves to the holographic CFTs where the correlator can be approximated by a single conformal block.
In the $x\rightarrow 1$ limit, the exchange operator is the identity and negativity is identically zero \cite{2013JSMTE..02..008C,2014JHEP...09..010K}. We are thus concerned with the $x\rightarrow 0$ limit where the exchanged operator in this block, $\sigma^2_{n_e}$, has a conformal dimension of the order of $c$; therefore, the explicit form is unknown.
Nevertheless, we can use the Zamolodchikov recursion relation \cite{Zamolodchikov1987,Zamolodchikov1984} to evaluate the conformal block numerically to arbitrarily high precision.

\begin{figure}
 \begin{center}
  \includegraphics[width=8.0cm,clip]{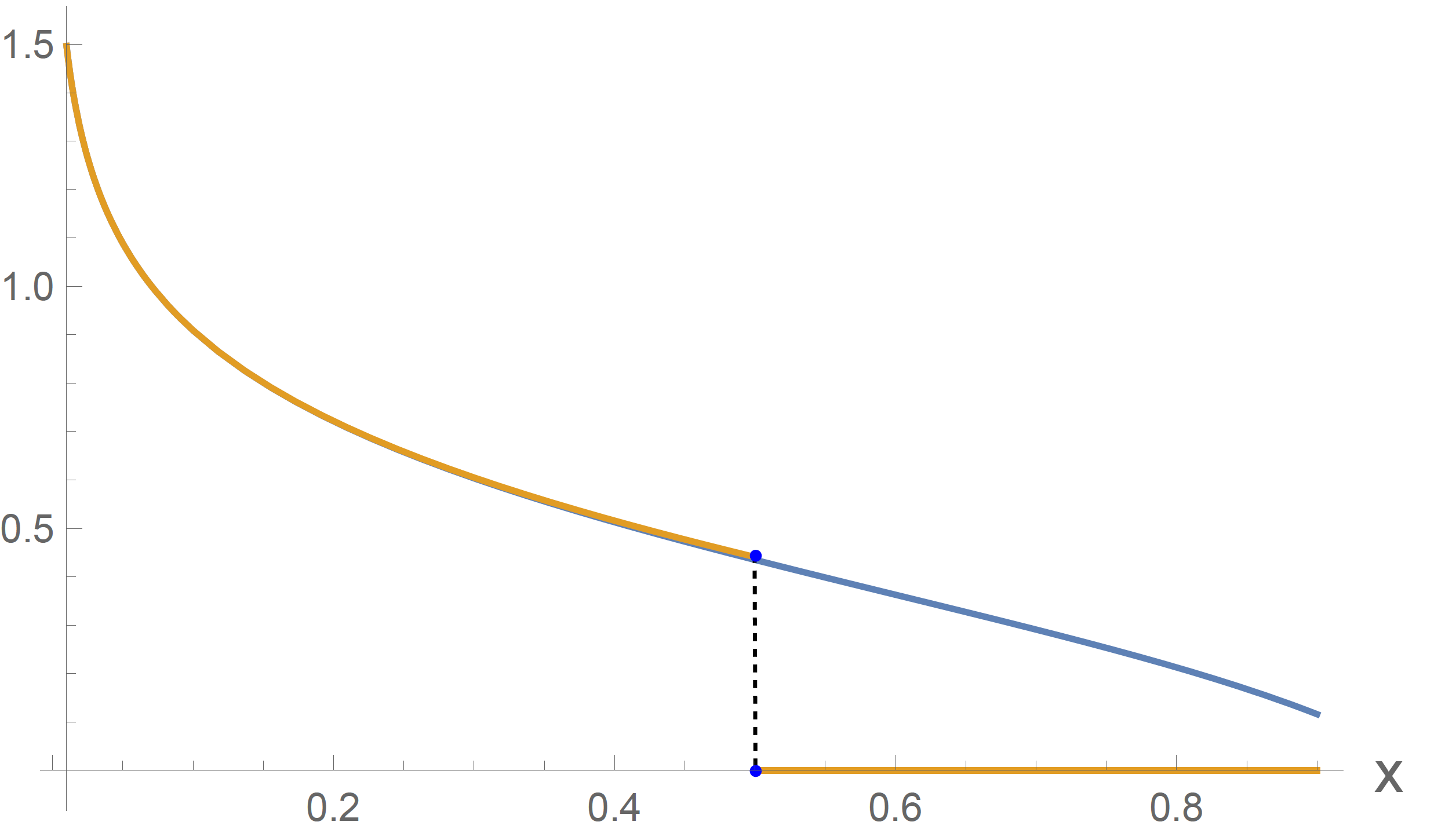}
 \end{center}
 \caption{The blue line is the negativity that comes from the Virasoro block computed to order $q^{500}$ using Zamolodchikov's recursion relation where $q$ is the elliptic nome. The yellow line shows the minimal entanglement wedge cross section (\ref{ew_disjoing}).
Here we set $c=10$ and $\e=10^{-2}$, and in this plot, we divide these quantities by $c$ to rescale.}
 \label{fig:numerical}
\end{figure}

On this background, the negativity should be compared to the entanglement wedge cross section, which is given by \cite{2018NatPh..14..573U}
\begin{equation}
\label{ew_disjoing}
E_W=
\left\{
\begin{array}{ll}
\displaystyle
\fr{c}{6} \log \fr{1+\s{1-x}}{1-\s{1-x}},
&
0 < x < 1/2,
\\
\\
0,
&
1/2 < x < 1.
\end{array}
\right.
\end{equation}
In Fig.~\ref{fig:numerical}, we show the entanglement wedge cross section and the negativity calculated by the Zamolodchikov recursion relation.
One can immediately find that the negativity perfectly matches the minimal entanglement wedge cross section. This result resolves the mysteries that arose when computing (\ref{LN_disjoint_corr}) using less direct approaches in Refs. \cite{2014JHEP...09..010K, PhysRevD.99.106014,2019arXiv190607639K}.

We progress to 
the negativity of a single interval 
of length $l$
at finite temperature $\beta^{-1}$,
which can be calculated by 
\footnote{The factor of $(2\epsilon)^{4h_{n_e}^{(2)}}$ has been largely ignored in the literature but is necessary. It comes from the OPE of the twist fields in the original six-point correlation function into double twist fields in the resulting four-point function.} \cite{2015JPhA...48a5006C}
\begin{equation}\label{eq:defE}
\begin{aligned}
\mathcal{E}&=\lim_{L\to\infty} \lim_{n_e \to 1} \log \Big(\braket{ \sigma_{n_e}(-L) \bar{\sigma}^2_{n_e}(-l) \sigma^2_{n_e}(0) \bar{\sigma}_{n_e}(L)}_\beta \Big) 
\\ 
&\qquad 
\times (2\epsilon)^{4h_{n_e}^{(2)}} \Big(C_{ \sigma_{n_e} \bar{\sigma}^2_{n_e} \sigma_{n_e}}   \Big) ^2  
\\
&=\fr{c}{2} \log \left({\fr{\b}{2\pi \e }\ex{\fr{\pi l}{\beta}  }} \right)
\\
&\quad 
+  
\lim_{L\to\infty} \lim_{n_e \to 1}\log\braket{ \sigma_{n_e}(\infty) \bar{\sigma}^2_{n_e}(1) \sigma^2_{n_e}(x,\bar{x}) \bar{\sigma}_{n_e}(0)   },
\end{aligned}
\end{equation}
where 
$\epsilon$ is an UV regulator, and
the cross ratio is given by
\begin{equation}
x \ar{L\to \infty} \ex{-\fr{2 \pi l}{\b}}.
\end{equation}
We assume the OPE coefficient can be fixed by equating the pure state limits of (\ref{LN_corr}) and (\ref{SR_corr})
\begin{equation}
\lim_{n_e \rightarrow 1} C_{ \sigma_{n_e} \bar{\sigma}^2_{n_e} \sigma_{n_e}} = 2^{c/2},
\end{equation}
though it is possible that the pure state limits are different from the pure state themselves.
This assumption only changes the final answer by a constant shift.

In the holographic CFT, this four-point function can be approximated by a single conformal block.
The dominant conformal block has two candidates.

(i) \textit{s} channel: In the semiclassical limit, the \textit{s} channel block is simplified because the only contribution to the intermediate state is the primary exchange \cite{2014JHEP...08..145F}. Thus, the approximated correlator is given by
\newsavebox{\boxpa}
\sbox{\boxpa}{\includegraphics[width=100pt]{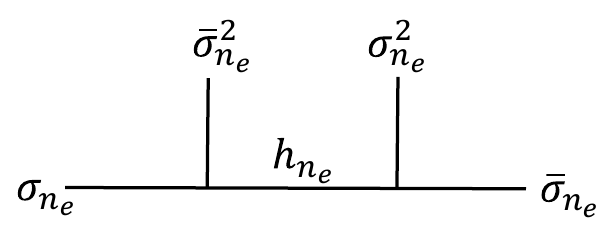}}
\newlength{\paw}
\settowidth{\paw}{\usebox{\boxpa}} 
\begin{equation}
\begin{aligned}
\abs{\parbox{\paw}{\usebox{\boxpa}} }^2  \pa{C_{ \sigma_{n_e} \bar{\sigma}^2_{n_e} \sigma_{n_e}}}^2 
\ar{n_e \to 1} \abs{x^{\fr{c}{8}}}^2 2^{-\fr{c}{2}}
.
\end{aligned}
\end{equation}
Substituting this into (\ref{eq:defE}), we obtain 
\begin{equation}
\ca{E}=\fr{c}{2} \log \fr{\b}{\pi \e}.
\end{equation}

(ii) \textit{t} channel: The \textit{t} channel block is just the \textit{HHLL} block \cite{2014JHEP...08..145F,2015JHEP...11..200F}, whose explicit form is given by
\begin{equation}\label{eq:HHLLblock}
\begin{aligned}
\ca{F}^{HH}_{LL}(0|1-z)  &= \fr{z^{2h_L}}{(1-z)^{2h_H}}  \ca{F}^{LL}_{HH}(0|1-z)    \\
&=  \fr{z^{h_L(\d+1)}}{  (1-z)^{2h_H} } \pa{ \fr{1-z^\d}{\d}}^{-2h_L},
\end{aligned}
\end{equation}
where $\d=\s{1-\fr{24}{c}h_H}$. 
Using this, we obtain
\newsavebox{\boxpb}
\sbox{\boxpb}{\includegraphics[width=100pt]{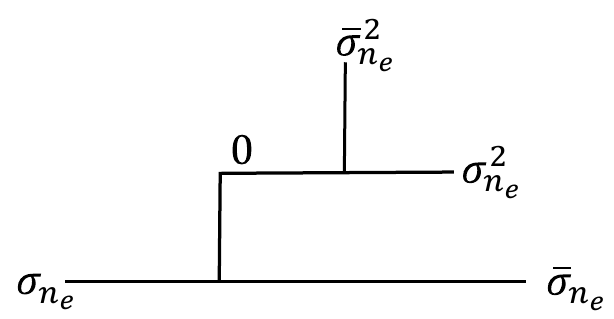}}
\newlength{\pbw}
\settowidth{\pbw}{\usebox{\boxpb}} 
\begin{equation}
\begin{aligned}
\abs{\parbox{\pbw}{\usebox{\boxpb}} }^2 
\ar{n_e \to 1} (1-x)^{\fr{c}{2}}.
\end{aligned}
\end{equation}
This leads to the negativity
\begin{equation}
\ca{E}=\fr{c}{2} \log \pa{\fr{\b}{\pi \e }  \sinh{\fr{\pi l}{\b}}   } . 
\end{equation}

Thus, we can conclude that the negativity at finite temperature is
\begin{equation}
\ca{E}=\min \BR{   \fr{c}{2} \log \fr{\b}{\pi \e},     \fr{c}{2} \log \pa{\fr{\b}{\pi \e }  \sinh{\fr{\pi l}{\b}}   }     }.
\end{equation}
This result perfectly matches the minimal entanglement wedge cross section in the Ba\~{n}ados Teitelboim Zanelli black hole geometry \cite{2018NatPh..14..573U} upon using (\ref{simple_LN}).


\textit{Discussion.---}In this Letter, we have derived the holographic dual of logarithmic negativity in AdS$_3$/CFT$_2$, confirming the conjecture from Ref.\ \cite{PhysRevD.99.106014}. There are several directions for future understanding. In particular, it is important to prove the holographic conjecture in higher dimensions. 3D gravity and 2D conformal field theory are quite special, so we are far from a complete derivation. Such a derivation may require a clever implementation of the gravitational replica trick for the partially transposed density matrix or a comparison of correlation functions of higher-dimensional twist operators for negativity and R\'enyi reflected entropy. Furthermore, the reflected entropy is not proven to (though it is believed to) be dual to the entanglement wedge cross section in generic dynamical settings. Because we have used the reflected entropy as a crutch in our derivation, the dynamical proposal for logarithmic negativity in AdS$_3$/CFT$_2$ is also still conjectural. Finally, it would be interesting to investigate quantum corrections to the holographic formula in the sense of Ref.~\cite{2013JHEP...11..074F}. General expectations and hints from error-correcting codes lead us to guess that the leading correction comes from the logarithmic negativity between the bulk fields on either side of the entanglement wedge cross section.

\acknowledgements
We thank Souvik Dutta, Tom Faulkner, and Tadashi Takayanagi for useful discussions. {We thank Yuhan Liu for discussions regarding the OPE coefficient assumption.} S.R. is supported by a Simons Investigator Grant from the Simons Foundation. Y.K. is supported by a JSPS fellowship. J.K.F. and S.R. thank the Yukawa Institute for Theoretical Physics (YITP-T-19-03) for hospitality during the completion of this work.

%


\begin{thebibliography}{35}%
\makeatletter
\providecommand \@ifxundefined [1]{%
 \@ifx{#1\undefined}
}%
\providecommand \@ifnum [1]{%
 \ifnum #1\expandafter \@firstoftwo
 \else \expandafter \@secondoftwo
 \fi
}%
\providecommand \@ifx [1]{%
 \ifx #1\expandafter \@firstoftwo
 \else \expandafter \@secondoftwo
 \fi
}%
\providecommand \natexlab [1]{#1}%
\providecommand \enquote  [1]{``#1''}%
\providecommand \bibnamefont  [1]{#1}%
\providecommand \bibfnamefont [1]{#1}%
\providecommand \citenamefont [1]{#1}%
\providecommand \href@noop [0]{\@secondoftwo}%
\providecommand \href [0]{\begingroup \@sanitize@url \@href}%
\providecommand \@href[1]{\@@startlink{#1}\@@href}%
\providecommand \@@href[1]{\endgroup#1\@@endlink}%
\providecommand \@sanitize@url [0]{\catcode `\\12\catcode `\$12\catcode
  `\&12\catcode `\#12\catcode `\^12\catcode `\_12\catcode `\%12\relax}%
\providecommand \@@startlink[1]{}%
\providecommand \@@endlink[0]{}%
\providecommand \url  [0]{\begingroup\@sanitize@url \@url }%
\providecommand \@url [1]{\endgroup\@href {#1}{\urlprefix }}%
\providecommand \urlprefix  [0]{URL }%
\providecommand \Eprint [0]{\href }%
\providecommand \doibase [0]{http://dx.doi.org/}%
\providecommand \selectlanguage [0]{\@gobble}%
\providecommand \bibinfo  [0]{\@secondoftwo}%
\providecommand \bibfield  [0]{\@secondoftwo}%
\providecommand \translation [1]{[#1]}%
\providecommand \BibitemOpen [0]{}%
\providecommand \bibitemStop [0]{}%
\providecommand \bibitemNoStop [0]{.\EOS\space}%
\providecommand \EOS [0]{\spacefactor3000\relax}%
\providecommand \BibitemShut  [1]{\csname bibitem#1\endcsname}%
\let\auto@bib@innerbib\@empty
\bibitem [{\citenamefont {{Ryu}}\ and\ \citenamefont
  {{Takayanagi}}(2006{\natexlab{a}})}]{2006PhRvL..96r1602R}%
  \BibitemOpen
  \bibfield  {author} {\bibinfo {author} {\bibfnamefont {S.}~\bibnamefont
  {{Ryu}}}\ and\ \bibinfo {author} {\bibfnamefont {T.}~\bibnamefont
  {{Takayanagi}}},\ }\href {\doibase 10.1103/PhysRevLett.96.181602} {\bibfield
  {journal} {\bibinfo  {journal} {\prl}\ }\textbf {\bibinfo {volume} {96}},\
  \bibinfo {eid} {181602} (\bibinfo {year} {2006}{\natexlab{a}})},\ \Eprint
  {http://arxiv.org/abs/hep-th/0603001} {arXiv:hep-th/0603001 [hep-th]}
  \BibitemShut {NoStop}%
\bibitem [{\citenamefont {{Ryu}}\ and\ \citenamefont
  {{Takayanagi}}(2006{\natexlab{b}})}]{2006JHEP...08..045R}%
  \BibitemOpen
  \bibfield  {author} {\bibinfo {author} {\bibfnamefont {S.}~\bibnamefont
  {{Ryu}}}\ and\ \bibinfo {author} {\bibfnamefont {T.}~\bibnamefont
  {{Takayanagi}}},\ }\href {\doibase 10.1088/1126-6708/2006/08/045} {\bibfield
  {journal} {\bibinfo  {journal} {Journal of High Energy Physics}\ }\textbf
  {\bibinfo {volume} {2006}},\ \bibinfo {eid} {045} (\bibinfo {year}
  {2006}{\natexlab{b}})},\ \Eprint {http://arxiv.org/abs/hep-th/0605073}
  {arXiv:hep-th/0605073 [hep-th]} \BibitemShut {NoStop}%
\bibitem [{\citenamefont {{Hubeny}}\ \emph {et~al.}(2007)\citenamefont
  {{Hubeny}}, \citenamefont {{Rangamani}},\ and\ \citenamefont
  {{Takayanagi}}}]{2007JHEP...07..062H}%
  \BibitemOpen
  \bibfield  {author} {\bibinfo {author} {\bibfnamefont {V.~E.}\ \bibnamefont
  {{Hubeny}}}, \bibinfo {author} {\bibfnamefont {M.}~\bibnamefont
  {{Rangamani}}}, \ and\ \bibinfo {author} {\bibfnamefont {T.}~\bibnamefont
  {{Takayanagi}}},\ }\href {\doibase 10.1088/1126-6708/2007/07/062} {\bibfield
  {journal} {\bibinfo  {journal} {Journal of High Energy Physics}\ }\textbf
  {\bibinfo {volume} {2007}},\ \bibinfo {eid} {062} (\bibinfo {year} {2007})},\
  \Eprint {http://arxiv.org/abs/0705.0016} {arXiv:0705.0016 [hep-th]}
  \BibitemShut {NoStop}%
\bibitem [{Note1()}]{Note1}%
  \BibitemOpen
  \bibinfo {note} {It is important to note that logarithmic negativity only
  provides an upper bound on the distillable entanglement.}\BibitemShut {Stop}%
\bibitem [{\citenamefont {Peres}(1996)}]{PhysRevLett.77.1413}%
  \BibitemOpen
  \bibfield  {author} {\bibinfo {author} {\bibfnamefont {A.}~\bibnamefont
  {Peres}},\ }\href {\doibase 10.1103/PhysRevLett.77.1413} {\bibfield
  {journal} {\bibinfo  {journal} {Phys. Rev. Lett.}\ }\textbf {\bibinfo
  {volume} {77}},\ \bibinfo {pages} {1413} (\bibinfo {year}
  {1996})}\BibitemShut {NoStop}%
\bibitem [{\citenamefont {{Horodecki}}\ \emph {et~al.}(1996)\citenamefont
  {{Horodecki}}, \citenamefont {{Horodecki}},\ and\ \citenamefont
  {{Horodecki}}}]{1996PhLA..223....1H}%
  \BibitemOpen
  \bibfield  {author} {\bibinfo {author} {\bibfnamefont {M.}~\bibnamefont
  {{Horodecki}}}, \bibinfo {author} {\bibfnamefont {P.}~\bibnamefont
  {{Horodecki}}}, \ and\ \bibinfo {author} {\bibfnamefont {R.}~\bibnamefont
  {{Horodecki}}},\ }\href {\doibase 10.1016/S0375-9601(96)00706-2} {\bibfield
  {journal} {\bibinfo  {journal} {Physics Letters A}\ }\textbf {\bibinfo
  {volume} {223}},\ \bibinfo {pages} {1} (\bibinfo {year} {1996})},\ \Eprint
  {http://arxiv.org/abs/quant-ph/9605038} {arXiv:quant-ph/9605038 [quant-ph]}
  \BibitemShut {NoStop}%
\bibitem [{\citenamefont {{Eisert}}\ and\ \citenamefont
  {{Plenio}}(1999)}]{1999JMOp...46..145E}%
  \BibitemOpen
  \bibfield  {author} {\bibinfo {author} {\bibfnamefont {J.}~\bibnamefont
  {{Eisert}}}\ and\ \bibinfo {author} {\bibfnamefont {M.~B.}\ \bibnamefont
  {{Plenio}}},\ }\href {\doibase 10.1080/09500349908231260} {\bibfield
  {journal} {\bibinfo  {journal} {Journal of Modern Optics}\ }\textbf {\bibinfo
  {volume} {46}},\ \bibinfo {pages} {145} (\bibinfo {year} {1999})},\ \Eprint
  {http://arxiv.org/abs/quant-ph/9807034} {arXiv:quant-ph/9807034 [quant-ph]}
  \BibitemShut {NoStop}%
\bibitem [{\citenamefont {{Simon}}(2000)}]{2000PhRvL..84.2726S}%
  \BibitemOpen
  \bibfield  {author} {\bibinfo {author} {\bibfnamefont {R.}~\bibnamefont
  {{Simon}}},\ }\href {\doibase 10.1103/PhysRevLett.84.2726} {\bibfield
  {journal} {\bibinfo  {journal} {\prl}\ }\textbf {\bibinfo {volume} {84}},\
  \bibinfo {pages} {2726} (\bibinfo {year} {2000})},\ \Eprint
  {http://arxiv.org/abs/quant-ph/9909044} {arXiv:quant-ph/9909044 [quant-ph]}
  \BibitemShut {NoStop}%
\bibitem [{\citenamefont {{Vidal}}\ and\ \citenamefont
  {{Werner}}(2002)}]{2002PhRvA..65c2314V}%
  \BibitemOpen
  \bibfield  {author} {\bibinfo {author} {\bibfnamefont {G.}~\bibnamefont
  {{Vidal}}}\ and\ \bibinfo {author} {\bibfnamefont {R.~F.}\ \bibnamefont
  {{Werner}}},\ }\href {\doibase 10.1103/PhysRevA.65.032314} {\bibfield
  {journal} {\bibinfo  {journal} {\pra}\ }\textbf {\bibinfo {volume} {65}},\
  \bibinfo {eid} {032314} (\bibinfo {year} {2002})},\ \Eprint
  {http://arxiv.org/abs/quant-ph/0102117} {arXiv:quant-ph/0102117 [quant-ph]}
  \BibitemShut {NoStop}%
\bibitem [{\citenamefont {{Plenio}}(2005)}]{2005PhRvL..95i0503P}%
  \BibitemOpen
  \bibfield  {author} {\bibinfo {author} {\bibfnamefont {M.~B.}\ \bibnamefont
  {{Plenio}}},\ }\href {\doibase 10.1103/PhysRevLett.95.090503} {\bibfield
  {journal} {\bibinfo  {journal} {\prl}\ }\textbf {\bibinfo {volume} {95}},\
  \bibinfo {eid} {090503} (\bibinfo {year} {2005})},\ \Eprint
  {http://arxiv.org/abs/quant-ph/0505071} {arXiv:quant-ph/0505071 [quant-ph]}
  \BibitemShut {NoStop}%
\bibitem [{\citenamefont {Kudler-Flam}\ and\ \citenamefont
  {Ryu}(2019)}]{PhysRevD.99.106014}%
  \BibitemOpen
  \bibfield  {author} {\bibinfo {author} {\bibfnamefont {J.}~\bibnamefont
  {Kudler-Flam}}\ and\ \bibinfo {author} {\bibfnamefont {S.}~\bibnamefont
  {Ryu}},\ }\href {\doibase 10.1103/PhysRevD.99.106014} {\bibfield  {journal}
  {\bibinfo  {journal} {Phys. Rev. D}\ }\textbf {\bibinfo {volume} {99}},\
  \bibinfo {pages} {106014} (\bibinfo {year} {2019})}\BibitemShut {NoStop}%
\bibitem [{\citenamefont {{Umemoto}}\ and\ \citenamefont
  {{Takayanagi}}(2018)}]{2018NatPh..14..573U}%
  \BibitemOpen
  \bibfield  {author} {\bibinfo {author} {\bibfnamefont {K.}~\bibnamefont
  {{Umemoto}}}\ and\ \bibinfo {author} {\bibfnamefont {T.}~\bibnamefont
  {{Takayanagi}}},\ }\href {\doibase 10.1038/s41567-018-0075-2} {\bibfield
  {journal} {\bibinfo  {journal} {Nature Physics}\ }\textbf {\bibinfo {volume}
  {14}},\ \bibinfo {pages} {573} (\bibinfo {year} {2018})},\ \Eprint
  {http://arxiv.org/abs/1708.09393} {arXiv:1708.09393 [hep-th]} \BibitemShut
  {NoStop}%
\bibitem [{\citenamefont {{Nguyen}}\ \emph {et~al.}(2018)\citenamefont
  {{Nguyen}}, \citenamefont {{Devakul}}, \citenamefont {{Halbasch}},
  \citenamefont {{Zaletel}},\ and\ \citenamefont
  {{Swingle}}}]{2018JHEP...01..098N}%
  \BibitemOpen
  \bibfield  {author} {\bibinfo {author} {\bibfnamefont {P.}~\bibnamefont
  {{Nguyen}}}, \bibinfo {author} {\bibfnamefont {T.}~\bibnamefont {{Devakul}}},
  \bibinfo {author} {\bibfnamefont {M.~G.}\ \bibnamefont {{Halbasch}}},
  \bibinfo {author} {\bibfnamefont {M.~P.}\ \bibnamefont {{Zaletel}}}, \ and\
  \bibinfo {author} {\bibfnamefont {B.}~\bibnamefont {{Swingle}}},\ }\href
  {\doibase 10.1007/JHEP01(2018)098} {\bibfield  {journal} {\bibinfo  {journal}
  {Journal of High Energy Physics}\ }\textbf {\bibinfo {volume} {2018}},\
  \bibinfo {eid} {98} (\bibinfo {year} {2018})},\ \Eprint
  {http://arxiv.org/abs/1709.07424} {arXiv:1709.07424 [hep-th]} \BibitemShut
  {NoStop}%
\bibitem [{\citenamefont {Tamaoka}(2019)}]{PhysRevLett.122.141601}%
  \BibitemOpen
  \bibfield  {author} {\bibinfo {author} {\bibfnamefont {K.}~\bibnamefont
  {Tamaoka}},\ }\href {\doibase 10.1103/PhysRevLett.122.141601} {\bibfield
  {journal} {\bibinfo  {journal} {Phys. Rev. Lett.}\ }\textbf {\bibinfo
  {volume} {122}},\ \bibinfo {pages} {141601} (\bibinfo {year}
  {2019})}\BibitemShut {NoStop}%
\bibitem [{\citenamefont {{Dutta}}\ and\ \citenamefont
  {{Faulkner}}(2019)}]{2019arXiv190500577D}%
  \BibitemOpen
  \bibfield  {author} {\bibinfo {author} {\bibfnamefont {S.}~\bibnamefont
  {{Dutta}}}\ and\ \bibinfo {author} {\bibfnamefont {T.}~\bibnamefont
  {{Faulkner}}},\ }\href@noop {} {\bibfield  {journal} {\bibinfo  {journal}
  {arXiv e-prints}\ ,\ \bibinfo {eid} {arXiv:1905.00577}} (\bibinfo {year}
  {2019})},\ \Eprint {http://arxiv.org/abs/1905.00577} {arXiv:1905.00577
  [hep-th]} \BibitemShut {NoStop}%
\bibitem [{\citenamefont {{Lewkowycz}}\ and\ \citenamefont
  {{Maldacena}}(2013)}]{2013JHEP...08..090L}%
  \BibitemOpen
  \bibfield  {author} {\bibinfo {author} {\bibfnamefont {A.}~\bibnamefont
  {{Lewkowycz}}}\ and\ \bibinfo {author} {\bibfnamefont {J.}~\bibnamefont
  {{Maldacena}}},\ }\href {\doibase 10.1007/JHEP08(2013)090} {\bibfield
  {journal} {\bibinfo  {journal} {Journal of High Energy Physics}\ }\textbf
  {\bibinfo {volume} {2013}},\ \bibinfo {eid} {90} (\bibinfo {year} {2013})},\
  \Eprint {http://arxiv.org/abs/1304.4926} {arXiv:1304.4926 [hep-th]}
  \BibitemShut {NoStop}%
\bibitem [{\citenamefont {{Kusuki}}\ and\ \citenamefont
  {{Tamaoka}}(2019{\natexlab{a}})}]{2019arXiv190706646K}%
  \BibitemOpen
  \bibfield  {author} {\bibinfo {author} {\bibfnamefont {Y.}~\bibnamefont
  {{Kusuki}}}\ and\ \bibinfo {author} {\bibfnamefont {K.}~\bibnamefont
  {{Tamaoka}}},\ }\href@noop {} {\bibfield  {journal} {\bibinfo  {journal}
  {arXiv e-prints}\ ,\ \bibinfo {eid} {arXiv:1907.06646}} (\bibinfo {year}
  {2019}{\natexlab{a}})},\ \Eprint {http://arxiv.org/abs/1907.06646}
  {arXiv:1907.06646 [hep-th]} \BibitemShut {NoStop}%
\bibitem [{\citenamefont {{Dong}}(2016)}]{2016NatCo...712472D}%
  \BibitemOpen
  \bibfield  {author} {\bibinfo {author} {\bibfnamefont {X.}~\bibnamefont
  {{Dong}}},\ }\href {\doibase 10.1038/ncomms12472} {\bibfield  {journal}
  {\bibinfo  {journal} {Nature Communications}\ }\textbf {\bibinfo {volume}
  {7}},\ \bibinfo {eid} {12472} (\bibinfo {year} {2016})},\ \Eprint
  {http://arxiv.org/abs/1601.06788} {arXiv:1601.06788 [hep-th]} \BibitemShut
  {NoStop}%
\bibitem [{Note2()}]{Note2}%
  \BibitemOpen
  \bibinfo {note} {In higher dimensions, the Engelhardt-Wall gluing procedure
  may be complicated by shockwaves associated with the conical singularities
  \cite {2019JHEP...05..160E}.}\BibitemShut {Stop}%
\bibitem [{Note3()}]{Note3}%
  \BibitemOpen
  \bibinfo {note} {Strictly speaking, this equality is only proven for the
  vacuum state \cite {2011JHEP...12..047H,2014JHEP...10..060R}. However, it is
  expected (and has passed multiple tests) that it holds whenever both the
  subsystem configuration and quantum state are spherically symmetric. We
  stress that it does not hold in generic situations.}\BibitemShut {Stop}%
\bibitem [{\citenamefont {{Calabrese}}\ \emph {et~al.}(2013)\citenamefont
  {{Calabrese}}, \citenamefont {{Cardy}},\ and\ \citenamefont
  {{Tonni}}}]{2013JSMTE..02..008C}%
  \BibitemOpen
  \bibfield  {author} {\bibinfo {author} {\bibfnamefont {P.}~\bibnamefont
  {{Calabrese}}}, \bibinfo {author} {\bibfnamefont {J.}~\bibnamefont
  {{Cardy}}}, \ and\ \bibinfo {author} {\bibfnamefont {E.}~\bibnamefont
  {{Tonni}}},\ }\href {\doibase 10.1088/1742-5468/2013/02/P02008} {\bibfield
  {journal} {\bibinfo  {journal} {Journal of Statistical Mechanics: Theory and
  Experiment}\ }\textbf {\bibinfo {volume} {2013}},\ \bibinfo {pages} {02008}
  (\bibinfo {year} {2013})},\ \Eprint {http://arxiv.org/abs/1210.5359}
  {arXiv:1210.5359 [cond-mat.stat-mech]} \BibitemShut {NoStop}%
\bibitem [{\citenamefont {{Calabrese}}\ \emph {et~al.}(2012)\citenamefont
  {{Calabrese}}, \citenamefont {{Cardy}},\ and\ \citenamefont
  {{Tonni}}}]{2012PhRvL.109m0502C}%
  \BibitemOpen
  \bibfield  {author} {\bibinfo {author} {\bibfnamefont {P.}~\bibnamefont
  {{Calabrese}}}, \bibinfo {author} {\bibfnamefont {J.}~\bibnamefont
  {{Cardy}}}, \ and\ \bibinfo {author} {\bibfnamefont {E.}~\bibnamefont
  {{Tonni}}},\ }\href {\doibase 10.1103/PhysRevLett.109.130502} {\bibfield
  {journal} {\bibinfo  {journal} {\prl}\ }\textbf {\bibinfo {volume} {109}},\
  \bibinfo {eid} {130502} (\bibinfo {year} {2012})},\ \Eprint
  {http://arxiv.org/abs/1206.3092} {arXiv:1206.3092 [cond-mat.stat-mech]}
  \BibitemShut {NoStop}%
\bibitem [{\citenamefont {{Kulaxizi}}\ \emph {et~al.}(2014)\citenamefont
  {{Kulaxizi}}, \citenamefont {{Parnachev}},\ and\ \citenamefont
  {{Policastro}}}]{2014JHEP...09..010K}%
  \BibitemOpen
  \bibfield  {author} {\bibinfo {author} {\bibfnamefont {M.}~\bibnamefont
  {{Kulaxizi}}}, \bibinfo {author} {\bibfnamefont {A.}~\bibnamefont
  {{Parnachev}}}, \ and\ \bibinfo {author} {\bibfnamefont {G.}~\bibnamefont
  {{Policastro}}},\ }\href {\doibase 10.1007/JHEP09(2014)010} {\bibfield
  {journal} {\bibinfo  {journal} {Journal of High Energy Physics}\ }\textbf
  {\bibinfo {volume} {2014}},\ \bibinfo {eid} {10} (\bibinfo {year} {2014})},\
  \Eprint {http://arxiv.org/abs/1407.0324} {arXiv:1407.0324 [hep-th]}
  \BibitemShut {NoStop}%
\bibitem [{\citenamefont {Zamolodchikov}(1987)}]{Zamolodchikov1987}%
  \BibitemOpen
  \bibfield  {author} {\bibinfo {author} {\bibfnamefont {A.~B.}\ \bibnamefont
  {Zamolodchikov}},\ }\href@noop {} {\bibfield  {journal} {\bibinfo  {journal}
  {Theoretical and Mathematical Physics}\ }\textbf {\bibinfo {volume} {73}},\
  \bibinfo {pages} {1088} (\bibinfo {year} {1987})}\BibitemShut {NoStop}%
\bibitem [{\citenamefont {Zamolodchikov}(1984)}]{Zamolodchikov1984}%
  \BibitemOpen
  \bibfield  {author} {\bibinfo {author} {\bibfnamefont {A.~B.}\ \bibnamefont
  {Zamolodchikov}},\ }\href {\doibase 10.1007/BF01214585} {\bibfield  {journal}
  {\bibinfo  {journal} {Commun. Math. Phys.}\ }\textbf {\bibinfo {volume}
  {96}},\ \bibinfo {pages} {419} (\bibinfo {year} {1984})}\BibitemShut
  {NoStop}%
\bibitem [{\citenamefont {{Kudler-Flam}}\ \emph {et~al.}(2019)\citenamefont
  {{Kudler-Flam}}, \citenamefont {{Nozaki}}, \citenamefont {{Ryu}},\ and\
  \citenamefont {{Tian Tan}}}]{2019arXiv190607639K}%
  \BibitemOpen
  \bibfield  {author} {\bibinfo {author} {\bibfnamefont {J.}~\bibnamefont
  {{Kudler-Flam}}}, \bibinfo {author} {\bibfnamefont {M.}~\bibnamefont
  {{Nozaki}}}, \bibinfo {author} {\bibfnamefont {S.}~\bibnamefont {{Ryu}}}, \
  and\ \bibinfo {author} {\bibfnamefont {M.}~\bibnamefont {{Tian Tan}}},\
  }\href@noop {} {\bibfield  {journal} {\bibinfo  {journal} {arXiv e-prints}\
  ,\ \bibinfo {eid} {arXiv:1906.07639}} (\bibinfo {year} {2019})},\ \Eprint
  {http://arxiv.org/abs/1906.07639} {arXiv:1906.07639 [hep-th]} \BibitemShut
  {NoStop}%
\bibitem [{Note4()}]{Note4}%
  \BibitemOpen
  \bibinfo {note} {The factor of $(2\epsilon )^{4h_{n_e}^{(2)}}$ has been
  largely ignored in the literature but is necessary. It comes from the OPE of
  the twist fields in the original six-point correlation function into double
  twist fields in the resulting four-point function.}\BibitemShut {Stop}%
\bibitem [{\citenamefont {{Calabrese}}\ \emph {et~al.}(2015)\citenamefont
  {{Calabrese}}, \citenamefont {{Cardy}},\ and\ \citenamefont
  {{Tonni}}}]{2015JPhA...48a5006C}%
  \BibitemOpen
  \bibfield  {author} {\bibinfo {author} {\bibfnamefont {P.}~\bibnamefont
  {{Calabrese}}}, \bibinfo {author} {\bibfnamefont {J.}~\bibnamefont
  {{Cardy}}}, \ and\ \bibinfo {author} {\bibfnamefont {E.}~\bibnamefont
  {{Tonni}}},\ }\href {\doibase 10.1088/1751-8113/48/1/015006} {\bibfield
  {journal} {\bibinfo  {journal} {Journal of Physics A Mathematical General}\
  }\textbf {\bibinfo {volume} {48}},\ \bibinfo {eid} {015006} (\bibinfo {year}
  {2015})},\ \Eprint {http://arxiv.org/abs/1408.3043} {arXiv:1408.3043
  [cond-mat.stat-mech]} \BibitemShut {NoStop}%
\bibitem [{\citenamefont {{Fitzpatrick}}\ \emph {et~al.}(2014)\citenamefont
  {{Fitzpatrick}}, \citenamefont {{Kaplan}},\ and\ \citenamefont
  {{Walters}}}]{2014JHEP...08..145F}%
  \BibitemOpen
  \bibfield  {author} {\bibinfo {author} {\bibfnamefont {A.~L.}\ \bibnamefont
  {{Fitzpatrick}}}, \bibinfo {author} {\bibfnamefont {J.}~\bibnamefont
  {{Kaplan}}}, \ and\ \bibinfo {author} {\bibfnamefont {M.~T.}\ \bibnamefont
  {{Walters}}},\ }\href {\doibase 10.1007/JHEP08(2014)145} {\bibfield
  {journal} {\bibinfo  {journal} {Journal of High Energy Physics}\ }\textbf
  {\bibinfo {volume} {2014}},\ \bibinfo {eid} {145} (\bibinfo {year} {2014})},\
  \Eprint {http://arxiv.org/abs/1403.6829} {arXiv:1403.6829 [hep-th]}
  \BibitemShut {NoStop}%
\bibitem [{\citenamefont {{Fitzpatrick}}\ \emph {et~al.}(2015)\citenamefont
  {{Fitzpatrick}}, \citenamefont {{Kaplan}},\ and\ \citenamefont
  {{Walters}}}]{2015JHEP...11..200F}%
  \BibitemOpen
  \bibfield  {author} {\bibinfo {author} {\bibfnamefont {A.~L.}\ \bibnamefont
  {{Fitzpatrick}}}, \bibinfo {author} {\bibfnamefont {J.}~\bibnamefont
  {{Kaplan}}}, \ and\ \bibinfo {author} {\bibfnamefont {M.~T.}\ \bibnamefont
  {{Walters}}},\ }\href {\doibase 10.1007/JHEP11(2015)200} {\bibfield
  {journal} {\bibinfo  {journal} {Journal of High Energy Physics}\ }\textbf
  {\bibinfo {volume} {2015}},\ \bibinfo {eid} {200} (\bibinfo {year} {2015})},\
  \Eprint {http://arxiv.org/abs/1501.05315} {arXiv:1501.05315 [hep-th]}
  \BibitemShut {NoStop}%
\bibitem [{\citenamefont {{Faulkner}}\ \emph {et~al.}(2013)\citenamefont
  {{Faulkner}}, \citenamefont {{Lewkowycz}},\ and\ \citenamefont
  {{Maldacena}}}]{2013JHEP...11..074F}%
  \BibitemOpen
  \bibfield  {author} {\bibinfo {author} {\bibfnamefont {T.}~\bibnamefont
  {{Faulkner}}}, \bibinfo {author} {\bibfnamefont {A.}~\bibnamefont
  {{Lewkowycz}}}, \ and\ \bibinfo {author} {\bibfnamefont {J.}~\bibnamefont
  {{Maldacena}}},\ }\href {\doibase 10.1007/JHEP11(2013)074} {\bibfield
  {journal} {\bibinfo  {journal} {Journal of High Energy Physics}\ }\textbf
  {\bibinfo {volume} {2013}},\ \bibinfo {eid} {74} (\bibinfo {year} {2013})},\
  \Eprint {http://arxiv.org/abs/1307.2892} {arXiv:1307.2892 [hep-th]}
  \BibitemShut {NoStop}%
\bibitem [{\citenamefont {{Engelhardt}}\ and\ \citenamefont
  {{Wall}}(2019)}]{2019JHEP...05..160E}%
  \BibitemOpen
  \bibfield  {author} {\bibinfo {author} {\bibfnamefont {N.}~\bibnamefont
  {{Engelhardt}}}\ and\ \bibinfo {author} {\bibfnamefont {A.~C.}\ \bibnamefont
  {{Wall}}},\ }\href {\doibase 10.1007/JHEP05(2019)160} {\bibfield  {journal}
  {\bibinfo  {journal} {Journal of High Energy Physics}\ }\textbf {\bibinfo
  {volume} {2019}},\ \bibinfo {eid} {160} (\bibinfo {year} {2019})},\ \Eprint
  {http://arxiv.org/abs/1806.01281} {arXiv:1806.01281 [hep-th]} \BibitemShut
  {NoStop}%
\bibitem [{\citenamefont {{Hung}}\ \emph {et~al.}(2011)\citenamefont {{Hung}},
  \citenamefont {{Myers}}, \citenamefont {{Smolkin}},\ and\ \citenamefont
  {{Yale}}}]{2011JHEP...12..047H}%
  \BibitemOpen
  \bibfield  {author} {\bibinfo {author} {\bibfnamefont {L.-Y.}\ \bibnamefont
  {{Hung}}}, \bibinfo {author} {\bibfnamefont {R.~C.}\ \bibnamefont {{Myers}}},
  \bibinfo {author} {\bibfnamefont {M.}~\bibnamefont {{Smolkin}}}, \ and\
  \bibinfo {author} {\bibfnamefont {A.}~\bibnamefont {{Yale}}},\ }\href
  {\doibase 10.1007/JHEP12(2011)047} {\bibfield  {journal} {\bibinfo  {journal}
  {Journal of High Energy Physics}\ }\textbf {\bibinfo {volume} {12}},\
  \bibinfo {eid} {47} (\bibinfo {year} {2011})},\ \Eprint
  {http://arxiv.org/abs/1110.1084} {arXiv:1110.1084 [hep-th]} \BibitemShut
  {NoStop}%
\bibitem [{\citenamefont {{Rangamani}}\ and\ \citenamefont
  {{Rota}}(2014)}]{2014JHEP...10..060R}%
  \BibitemOpen
  \bibfield  {author} {\bibinfo {author} {\bibfnamefont {M.}~\bibnamefont
  {{Rangamani}}}\ and\ \bibinfo {author} {\bibfnamefont {M.}~\bibnamefont
  {{Rota}}},\ }\href {\doibase 10.1007/JHEP10(2014)060} {\bibfield  {journal}
  {\bibinfo  {journal} {Journal of High Energy Physics}\ }\textbf {\bibinfo
  {volume} {2014}},\ \bibinfo {eid} {60} (\bibinfo {year} {2014})},\ \Eprint
  {http://arxiv.org/abs/1406.6989} {arXiv:1406.6989 [hep-th]} \BibitemShut
  {NoStop}%
\bibitem [{\citenamefont {{Kusuki}}\ and\ \citenamefont
  {{Tamaoka}}(2019{\natexlab{b}})}]{2019arXiv190906790K}%
  \BibitemOpen
  \bibfield  {author} {\bibinfo {author} {\bibfnamefont {Y.}~\bibnamefont
  {{Kusuki}}}\ and\ \bibinfo {author} {\bibfnamefont {K.}~\bibnamefont
  {{Tamaoka}}},\ }\href@noop {} {\bibfield  {journal} {\bibinfo  {journal}
  {arXiv e-prints}\ ,\ \bibinfo {eid} {arXiv:1909.06790}} (\bibinfo {year}
  {2019}{\natexlab{b}})},\ \Eprint {http://arxiv.org/abs/1909.06790}
  {arXiv:1909.06790 [hep-th]} \BibitemShut {NoStop}%
\end{thebibliography}


\onecolumngrid
\section*{Supplemental Material}
\beginsupplement

We consider normalized states created by generic operator insertions
\begin{align}
    \ket{\psi} = \frac{\prod_i^{n_{\mathcal{O}}}\mathcal{O}_i (x_i)\ket{0}}{\left \langle \prod_i^{n_{\mathcal{O}}} \left[\mathcal{O}_i(x_i)\right]^{\dagger}\mathcal{O}_i(x_i)\right\rangle^{1/2}}
\end{align}
The negativity is then computed as
\begin{align}
    \mathcal{E} = &\lim_{n_e \rightarrow 1} \log \frac{ \left<\prod_{i}^{n_A}\left( \sigma_{n_e}(u_i) \bar{\sigma}_{n_e}(v_i)\right) \prod_{i}^{n_B}\left( \bar{\sigma}_{n_e}(w_i) \sigma_{n_e}(y_i)\right)  \prod_i^{n_O}\left[\mathcal{O}^{\otimes n_e}_i(x_i)\right]^{\dagger}\mathcal{O}^{\otimes n_e}_i(x_i)\right>_{CFT^{\otimes n_e}}}
    {\left(\left \langle \prod_i^{n_{\mathcal{O}}} \left[\mathcal{O}_i(x_i)\right]^{\dagger}\mathcal{O}_i(x_i)\right\rangle \right)^{n_e}}.
    \label{LN_full_corr}
\end{align}
This should be compared to the R\'enyi reflected entropy which has a different normalization due to the canonical purification procedure and computed by the following path integral (see Ref.~\cite{2019arXiv190500577D} for further details)
\begin{align}
    S_R^{(n)} = \frac{1}{1-n}\lim_{m_e \rightarrow 1}\log \frac{\mathcal{Z}_{n,m_e}}{\left(\mathcal{Z}_{1,m_e}\right)^n},
\end{align}
where the replica partition function is defined by
\begin{align}
\mathcal{Z}_{n,m_e}
\equiv
 \left<\prod_{i}^{n_A}( \sigma_{g^{\ }_A}(u_i) \sigma_{g_A^{-1}}(v_i))
    \prod_{i}^{n_B}( {\sigma}_{g^{\ }_B}(w_i) \sigma_{g_B^{-1}}(y_i))  \prod_i^{n_O}[\mathcal{O}^{\otimes m_en}_i(x_i)]^{\dagger}\mathcal{O}^{\otimes m_e n}_i(x_i)
 \right>_{{\it CFT}^{\otimes m_en}}
 .
\end{align}
We use the notation $m_e$ to remind the reader that we are taking the limit from even integers to one. 
In  the following, we will show the negativity matches a half of the R\'enyi reflected entropy at index 1/2,
\begin{align}
    \frac{S_R^{(1/2)}}{2} = \lim_{m_e \rightarrow 1}\log \frac{\mathcal{Z}_{1/2,m_e}}{\left(\mathcal{Z}_{1,m_e}\right)^{1/2}}.
     \label{SR_full_corr}
\end{align}
These correlation functions appear, for example, when studying the reflected entropy following a local quantum quench or in a heavy state \cite{2019arXiv190706646K}.
There is a subtlety that we must discuss regarding the structure of the primary operators in the replica theory. For the negativity, the operator $\mathcal{O}^{\otimes n_e}$ has the tensor structure
\begin{align}
    \mathcal{O}^{\otimes n_e} = \mathcal{O}_{(1)}^{\otimes n_e/2} \otimes \mathcal{O}_{(2)}^{\otimes n_e/2},
\end{align}
where the subscript $1 (2)$ implies that the operator acts like the local primary operator $\mathcal{O}$ on the odd (even) numbered sheets and the identity on the even (odd) sheets. Analogously, this is how the double twist field (for $n_e$) decomposes in general
\begin{align}
    \sigma_{n_e}^2 = \sigma_{n_e/2}^{(1)} \otimes \sigma_{n_e/2}^{(2)}.
\end{align}
For the reflected entropy, the decomposition is 
\begin{align}
    \mathcal{O}^{\otimes m_e n} = \mathcal{O}^{\otimes n}_{(0)} \otimes \dots \otimes \mathcal{O}^{\otimes n}_{(m_e/2)} \otimes \dots
\end{align} 
where the subscript labels the replica copy in the $m_e$ direction where the operator acts.
The key point is that in the $m_e\rightarrow 1$ limit, the operator $\mathcal{O}^{\otimes m_en}$ does not reduce to $\mathcal{O}^{\otimes n}$ as one would naively expect but rather the square
\begin{align}
    \lim_{m_e \rightarrow 1} \mathcal{O}^{\otimes m_en} = \mathcal{O}^{\otimes n}_{(0)} \otimes \mathcal{O}^{\otimes n}_{(1/2)}.
\end{align}
Similarly, this is how the twist operators from Ref.~\cite{2019arXiv190500577D} reduce in the limit
\begin{align}
   \lim_{m_e \rightarrow 1}  \sigma_{g_A^{-1} g_B} = \sigma_n^{(0)} \otimes \sigma_n^{(1/2)}.
\end{align}
As a result from this squaring, we obtain the squared correlation function,
\begin{align}
\lim_{m_e \rightarrow 1} \mathcal{Z}_{1,m_e}
=\left \langle \prod_i^{n_{\mathcal{O}}} \left[\mathcal{O}_i(x_i)\right]^{\dagger}\mathcal{O}_i(x_i)\right\rangle^{2}
\end{align}
One can find that substituting this into the denominator of (\ref{SR_full_corr}) completely reproduces the denominator of (\ref{LN_full_corr}).
Note that this is crucial for the R\'enyi reflected entropy to reduce to twice the R\'enyi entropy for pure states, which is a true statement for any quantum state. 
We also stress that these operators do not interact with one another, just as the operators for negativity on opposite parity sheets did not interact.
This leads to a decoupling of the components of the conformal blocks e.g.
\begin{align}
    \bra{\sigma_{n_e/2}^{(1)} \otimes \sigma_{n_e/2}^{(2)}}\mathcal{O}^{\otimes n_e/2}_{(1)} \otimes \mathcal{O}^{\otimes n_e/2}_{(2)}\ket{\sigma_{n_e/2}^{(1)} \otimes \sigma_{n_e/2}^{(2)}} &= \bra{\sigma_{n_e/2}^{(1)}}\mathcal{O}^{\otimes n_e/2}_{(1)} \ket{\sigma_{n_e/2}^{(1)} }\bra{\sigma_{n_e/2}^{(2)}}\mathcal{O}^{\otimes n_e/2}_{(2)} \ket{\sigma_{n_e/2}^{(2)} },
    \\ \nonumber \\
    \bra{\sigma_n^{(0)} \otimes \sigma_{n}^{(m_e/2)}}\mathcal{O}^{\otimes n}_{(0)} \otimes \mathcal{O}^{\otimes n}_{(m_e/2)}\ket{\sigma_n^{(0)} \otimes \sigma_{n}^{(m_e/2)}} &= \bra{\sigma_n^{(0)}}\mathcal{O}^{\otimes n}_{(0)} \ket{\sigma_n^{(0)} }\bra{\sigma_n^{(m_e/2)}}\mathcal{O}^{\otimes n}_{(m_e/2)} \ket{\sigma_n^{(m_e/2)} }.
\end{align}
Then, in the appropriate $n_e \rightarrow 1, n\rightarrow 1/2, m_e \rightarrow 1, c\rightarrow \infty$ limit, both the numerators of (\ref{LN_full_corr}) and (\ref{SR_full_corr}) are described by precisely the same conformal blocks, hence confirming (13) of the main text for generic states in AdS$_3$/CFT$_2$. This decoupling is further justified by calculations of the reflected entropy in holographic and rational CFTs that are consistent with known facts \cite{2019arXiv190706646K,2019arXiv190906790K}. 

\end{document}